# ADASYN－Random Forest Based Intrusion Detection Model

ZHEWEI CHEN

College of Mathematics and Computer Science, Zhejiang Normal University, Jinhua, China

czw990928@gmail.com

WENWEN YU

School of Computing and Data Engineering, NingboTech University, Ningbo, China

yww.1999@outlook.com

LINYUE ZHOU

College of Mathematics and Computer Science, Zhejiang Normal University, Jinhua, China

zhoulinyue73@gmail.com

Abstract. Intrusion detection has been a key topic in the field of cyber security, and the common network threats nowadays have the characteristics of varieties and variation. Considering the serious imbalance of intrusion detection datasets will result in low classification performance on attack behaviors of small sample size and difficulty to detect network attacks accurately and efficiently, using Adaptive Synthetic Sampling (ADASYN) method to balance datasets was proposed in this paper. In addition, Random Forest algorithm was used to train intrusion detection classifiers. Through the comparative experiment of Intrusion detection on CICIDS 2017 dataset, it is found that ADASYN with Random Forest performs better. Based on the experimental results, the improvement of precision, recall, F1 scores and AUC values after ADASYN is then analyzed. Experiments show that the proposed method can be applied to intrusion detection with large data, and can effectively improve the classification accuracy of network attack behaviors. Compared with traditional machine learning models, it has better performance, generalization ability and robustness.

## 1 INTRODUCTION

With the development of computer technology, as well as the popularity of the Internet, cyber security is facing severe challenges. Criminals use system and network vulnerabilities to steal users' private information or launch network attacks in order to seize illegal profit. In recent years, a variety of security incidents and cyber-attacks occur frequently: In June 2020, more than 200 public security and legal departments in the United States were stolen 296GB of sensitive data files due to hacker intrusion; In April 2020, the government website of North Rhine Westphalia in Germany suffered phishing attacks and lost tens of millions of euros. Under these circumstances, how to effectively protect cyber security has become an urgent problem.

In order to prevent organizations and individuals from being attacked by hackers, network intrusion detection technology[1] plays an important role in ensuring information security, and is considered as one of the most promising methods to protect the network from various attacks. It takes machine learning algorithm as the core, and can quickly and effectively identify the type of sample data. However, it is not a simple task to classify network attacks. Due to the diversity and variability of network attacks, many existing classification methods cannot cope with the changing attacks and the emergence of new attacks. At the same time, the classification models widely used in intrusion detection system include Bayesian network, Logistic Regression[2], Support Vector Machine[3], Decision Trees[4] and so on. Although these methods have been applied to some extent, there is still a lot of room for improvement due to the existence of irrelevant or redundant features in the dataset and its class imbalance.

In this paper, improved Random Forest based on ADASYN[5] (ADASYN-Random Forest algorithm) is used to deal with the problem of class imbalance and applied innovatively to classify and detect network attack behaviors. This paper merges eight datasets in CICIDS 2017 to simulate benign dataflow and latest common attacks in real network traffic. After original datasets are preprocessed and divided into training sets and testing sets, three different sampling strategies: Random Under-Sampling, SMOTE and ADASYN are separately performed on the training sets to compare the performance on unbalanced datasets. The training is based on random forests, and then the obtained model is tested.

## 2 RELATED WORK

Intrusion Detection System[1] and Threat Hunting[6] are similar in principle and are usually equivalent to classification problems. They divide network behaviors into different categories through machine learning algorithms to distinguish between network attacks and benign network behaviors. Address the problems of intrusion detection, scholars at home and abroad have done a variety of research on network traffic anomalies. The main framework of these researches is detection algorithm based on the features, statistics, data mining and machine learning[7].

Zhang H[8] uses deep learning technology for network attack detection and proposes a new intrusion detection system based on de-noising automatic encoder[9]. It uses weighted loss function to reduce the dimension of feature set, and employs Multilayer Perceptron as classifier. However, the parameters of Multilayer Perceptron are difficult to adjust, and it is easy to fall into the problem of over fitting as the number of hidden layers increases; Gou J[7] uses Random Forests technology for network intrusion detection, and proposes two random noise reduction methods, which obtain high accuracy, but no further refining study on how to subdivide different types of network attacks. In addition, Jing H[10] uses the BP neural network algorithm to study the intrusion detection in the application layer, but this method has the defect of low efficiency in large-scale website intrusion detection; Alkasassbeh M[11] conducts intrusion detection tests on KDD Cup 99 using classical machine learning algorithms such as Decision Tree, Multilayer Perceptron and Bayesian Network, but generalization of single learner models is insufficient.

The above classification methods usually construct intrusion detection models based on the chosen metrics, which often ignore the imbalance and other statistical characteristics of the dataset, resulting in poor predictive performance of the trained models, such as low precision and recall rates of the models[12], and insufficient generalization ability.

This paper optimizes the random forest algorithm, establishes a network intrusion detection model based on ADASYN-Random Forest with high generalization, efficiency and accuracy, and compares it with various machine learning algorithm models in these three aspects: Precision, Recall, F1-Score and AUC to show the superiority of this algorithm.

### 2.1 Selection of Machine Learning Algorithms

In this paper, CICIDS 2017[13] intrusion detection evaluation dataset is used to compare the detection performance of several common machine learning algorithms using 10-fold cross-validation method. These machine learning algorithms are Naive Bayesian[14], Logistic Regression, KNN, CART and Random Forest[15]. The dataset used to build these models is consistent with the preprocessed but unsampled dataset used in Fourth Part experiment. The evaluation of detection is shown in Table 1 below.

Table 1 Classification Performance of Several Machine Learning Algorithms

| Algorithm | Macro-Average F1 |
| --- | --- |
| Naive Bayes | 40.38% |
| Logistic Regression | 46.05% |
| K-Nearest Neighbor | 85.97% |
| CART | 90.14% |
| Random Forest | 91.38% |

From Table 1, it can be seen that the detection performance of machine learning algorithm based on statistical analysis is poor, the Macro-Average F1 of Naive Bayes and Logistic Regression are 40.38% and 46.05%, respectively. In addition, the performance of single learner machine learning algorithms is very close, which stand at 85.97% for KNN and 90.14% for CART. Whereas, the classification performance of random forest is overall higher than machine learning algorithms and reach 91.38%. Therefore, random forest algorithm was chosen to train the classification model of intrusion detection.

### 2.2 Random Under-Sampling

The under-sampling method improves the classification accuracy of minority class by reducing the number of samples in majority class. Random Under-Sampling[16] (RUS) is the simplest under sampling method, which can balance the number of samples by randomly selecting a part of samples belonging to majority class and deleting them.

Random Under-Sampling can effectively shorten the training time of the model, and multiple random under sampling can be used to reduce the error caused by single random under sampling. However, randomly discarding samples may remove potentially useful information from majority classes, resulting in reduced classifier performance and unclear decision boundary. Moreover, the parameters of sampling strategy in Random Under-Sampling are very difficult to adjust. In this paper, Random Under-Sampling is applied as a comparative experiment with other Oversampling methods.

### 2.3 ADASYN Oversampling

In the process of model training, the quality of the dataset will also affect the prediction performance of the model. Data preprocessing is a necessary step before training, which involves oversampling[16] the original unbalanced data samples. ADASYN[5] (Adaptive Synthetic Sampling) was employed in this research. The algorithm gives different weights to different minority samples and uses some mechanism to automatically determine how many composite samples each minority sample needs to produce in order to achieve the goal of data balance. ADASYN is not like SMOTE[17], which synthesizes the same number of samples for each minority sample.

### 2.4 Random Forest Algorithm

As an ensemble learning classification method, Random Forest is a combination of Bagging and Random Subspace algorithm. It has the characteristics of high accuracy, fast training speed and strong generalization ability. Moreover, it can effectively prevent the over-fitting problem of the model. Random Forest algorithm first resamples the training data of each decision tree by Bootstrap[7], and then completes the modeling process by the decision tree. After that, the final prediction result is obtained based on the Voting algorithm by synthesizing the prediction of different decision trees in the model.

In this research, Gini Coefficient[18] was selected as the evaluation metric of single decision tree splitting in Random Forest model. The reason is it can reflect the proportion relationship of all kinds of samples and the proportion change of different kinds of samples, which is convenient for analysis and processing.

### 3 ADASYN - RANDOM FOREST BASED INTRUSION DETECTION MODEL

The unbalance of training dataset will result in the unbalance of training data of each tree extracted by the random forest algorithm in the first random process[15], which makes the decision tree unable to get good classification results on minority class samples, and will ultimately result in the random forest algorithm having a high accuracy on the classification or prediction results on the majority class samples, but not on the minority class samples. Traditional random forest algorithms do not perform well on unbalanced datasets. Therefore, ADASYN was used to oversample intrusion detection datasets for training random forests, and synthesizes some minority attacking behaviors to balance the data. This can make up for the deficiency of random forest algorithms in dealing with unbalanced data and improve the prediction ability of models. The framework of the model proposed in this paper is shown in Figure 1.

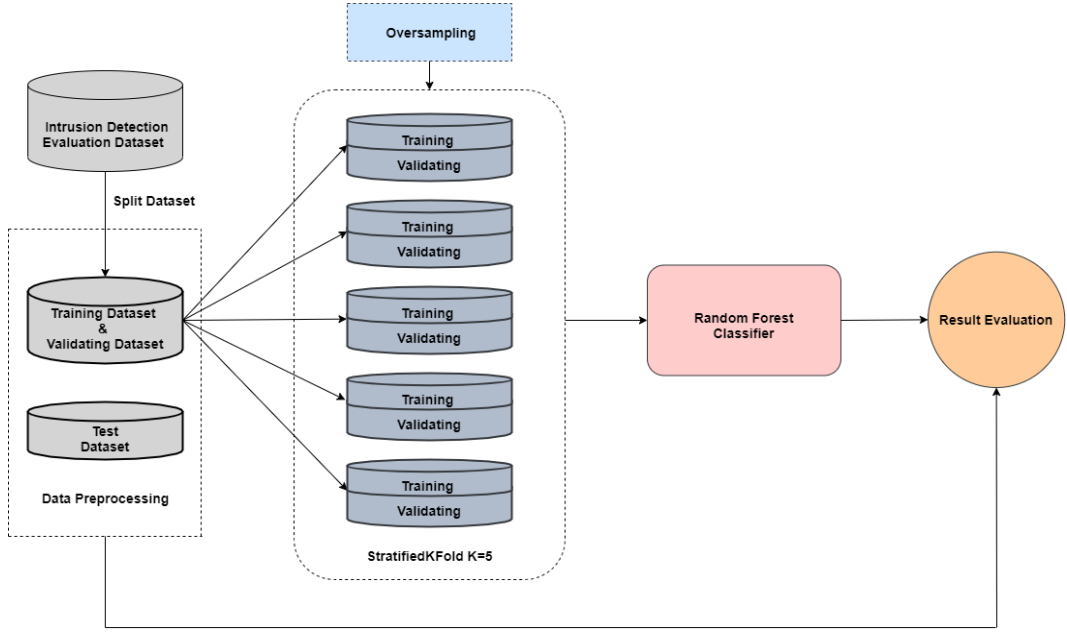

Figure 1. Framework of ADASYN-Random Forest based Intrusion Detection Predictor

Firstly, the original dataset is cleaned, then standardized, and divided into training set (80%) and test set (20%), that is training dataset ($X_T$) and prediction dataset ($X_P$). Then the training dataset is divided by StratifiedKFold cross validation. Meanwhile, the validation dataset ($X_V$) is separated. The training dataset is respectively oversampled by ADASYN with different sample strategy parameters. Subsequently, Random Forest Classifier is trained using the training dataset, and the optimal strategy parameters are determined using the validation dataset. Finally, both training and validation dataset are oversampled using optimal ADASYN method to train the classifier, the acquired model is tested by prediction dataset.

The pseudo code of ADASYN Oversampling is shown below:

Equation 1 ADASYN

| Algorithm 1: ADASYN |
|---|
| Input: Training dataset $X_T$, Hyper parameter $\beta \in [0,1]$, $K=5$ <br> The $i$ th sample in the minority sample $x_i$ ($i = 1, 2, 3, \cdots, m_s$), <br> A random minority sample $x_{zi}$ in K-nearest neighbors of $x_i$. |
| Output: Synthetic minority samples $s_i$, Oversampled training dataset $X_{ADASYN}$ |
| 1   Calculate the number of majority samples $m_l$ and the number of minority samples $m_s$ in Training dataset $X_T$ <br> 2   According to the formula $G = (m_l - m_s) \times \beta$ calculates the number of samples to be synthesized for minority class <br> 3   **For each example $x_i \in$ minorityclass:** <br> 4      Calculate $\Delta i$     //the number of majority samples in K-nearest neighbors of minority Sample $x_i$ <br> 5      Calculate $r_i = \Delta i/K$     //the ratio of majority samples in K-nearest neighbors of minority Sample $x_i$ <br> 6      Standardize $r_i$ through the formula $\hat{r}_i = r_i / \sum_{i=1}^{m_s} r_i$ <br> 7      Calculate $g_i = \hat{r}_i \times G$     //the number of new samples to be generated for each minority $x_i$ <br> 8      **Do the loop from 1 to $g_i$** <br> 9         Using the formula $s_i = x_i + (x_{zi} - x_i) \times \gamma$ to synthesize data samples     //$\gamma$ is a random number: $\gamma \in [0,1]$ <br> 10     **End** <br> 11  **End** <br> 12  **Return** Oversampled training datasets $X_{ADASYN}$ |

In algorithm 1, after inputting the training dataset and other parameters, ADASYN will automatically calculate the number of samples to be synthesized for different minority samples, and then generated samples according to SMOTE algorithm ($s_i = x_i + (x_{zi} - x_i) \times \gamma$). Since SMOTE algorithm treats all the minority samples equally and does not consider the class information of the nearest neighbor samples, the synthesized samples often produce noise or aliasing, which leads to poor classification effect; ADASYN algorithm uses density distribution as the criterion to generate more synthesis samples in the feature space region with low instance density, and generates more synthesis samples in the feature space region with high instance density. Therefore, in the case of extremely uneven distribution of network intrusion detection datasets, ADASYN can synthesize more effective minority samples and reduce the noise as much as possible.

In this research, Classification and Regression Trees (CART) were used to construct the decision trees of Random forest algorithm. As a result, Gini Coefficient is used as the criteria. It is used to measure the impurity of CART algorithm when splitting. For node $t$ in decision tree, Gini Coefficient is calculated as follows[7]:

$$Gini(t) = 1 - \sum_k [p(c_k|t)]^2 \qquad (1)$$

In the formula, the Gini Coefficient is the difference between 1 and the sum of the squares of the probabilities of class $c_k$, which represents the probability that a randomly selected sample in the sample set is misclassified. Generally speaking, the smaller the Gini index, the less likely the selected samples in the set will be misclassified, that is, the higher the purity of the set, and vice versa, the less pure the set. Compared with Entropy, this Gini Coefficient can better reflect the scale relationship of attack behavior in intrusion detection dataset and the scale change of different types of attack behavior[18], which is convenient for analysis and processing. After all the decision trees are generated, the Random Forest algorithm uses the majority voting method to classify and judge the categories of network dataflow based on the classification results of each CART.

The algorithm diagram for Random Forests is shown in Figure 2:

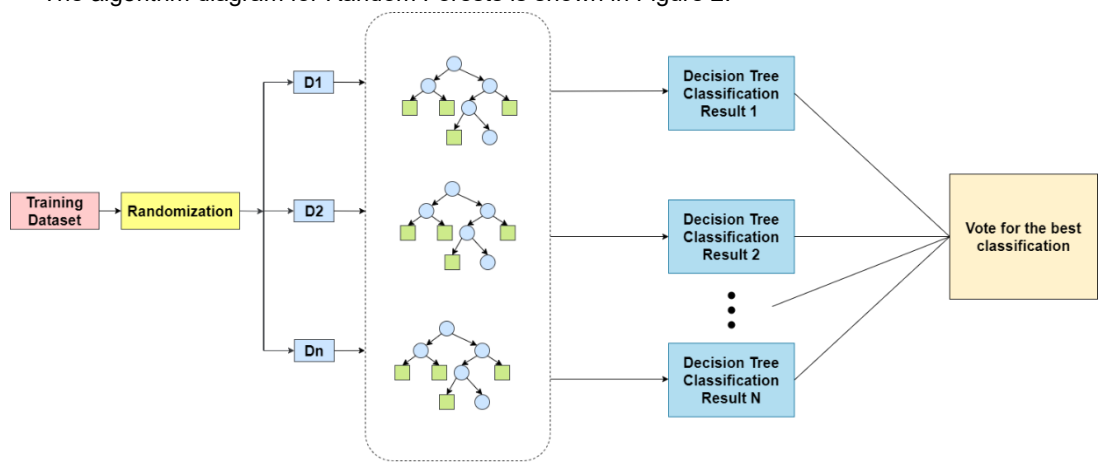

Figure 2. Framework of Random Forest Algorithm

The pseudo code of Random Forest is shown below:

Equation 2 Random Forest

| Algorithm 2: Random Forest |
| --- |
| Input:      Training dataset $X_T$, Number of decision trees $n\_estimators$, Maximum depth of decision tree $max\_depth$ Hyper parameter $k$, Feature selection criteria |
| Output:    Random forest classifier model, Classification results |
| 1 **For i = 1:** $n\_estimators$ |
| 2     Use Bootstrap method to select and give each tree a N-sized dataset          //Selection with replacement |
| 3     $k$ features are randomly selected at the nodes, the best features are compared and selected to divide the data set          // All features are selected in this paper |
| 4     Recursively generates decision trees, each with a maximum depth of $max\_depth$ with no pruning          // Feature selection criteria using Gini Coefficient |
| 5 **End** |
| 6 Calculate the probability that unknown sample $X$ is classified as $C$ in the test set: $P(C\|X) = (1/n\_estimators) \cdot \sum_{j=1}^{n\_estimators} h_j(C\|X)$ |
| 7 Classification by majority voting $C \leftarrow \arg\max P(C\|X)$, And calculate the classification error $OOB\ error$ |
| 8 **Return** Random forest classifier model, Classification results |

## 4 EXPERIMENTAL VERIFICATION

This part describes the basic setup of the experiment, and describes the whole process of the experiment, including dataset, data preprocessing, performance evaluation metrics and result analysis. This experiment was carried out in Windows 10 OS with i5-4590U CPU @3.30GHz and 8GB memory. We implement the methods mentioned in Section 3 with the Jupyter Lab platform in Anaconda.

### 4.1 Dataset

In this research, CICIDS 2017 intrusion detection evaluation dataset[13] was selected as the experimental dataset. Dataflow features such as Destination Port and Flow Duration is extracted from the network traffic. The dataset contains benign dataflows and latest common attacks, similar to real world data (PCAP), and its realistic background traffic has good simulation effect. In order to restore the real situation and simulate the real situation of intrusion detection, eight datasets with different kinds of network behaviors in CICIDS 2017 are merged in this paper. The following table shows the merged original datasets.

Table 2 Summary of Intrusion Detection Datasets

| Category | Attribute | Number |
| --- | --- | --- |
| BENIGN | 78 | 2273097 |
| DoS Hulk | 78 | 231073 |
| Port Scan | 78 | 158930 |
| DDoS | 78 | 128027 |
| Dos Golden Eye | 78 | 10293 |
| FTP-Patator | 78 | 7938 |
| SSH-Patator | 78 | 5897 |
| DoS slow loris | 78 | 5796 |
| DoS Slow http test | 78 | 5499 |
| Botnet | 78 | 1966 |
| Web Attack Bructe Force | 78 | 1507 |
| Web Attack XSS | 78 | 652 |
| Infiltration | 78 | 36 |
| Web Attack SQL Injection | 78 | 21 |
| Heartbleed | 78 | 11 |

### 4.2 Data Preprocessing

Considering that there are invalid or redundant attributes in the dataset (such as most of the values are the same for one specific attribute), these valid sample attributes are selected and similar classes with low instances are merged. What is more, considering that high correlation between features will affect the interpretability of the model and waste computing resources, the features with correlation coefficient higher than 0.95 were excluded. The datasets after preprocessing are shown in Table 3.

Table 3 Datasets after Preprocessing

| Category | Attribute | Number |
|---|---|---|
| BENIGN | 47 | 2072444 |
| DoS Attack | 47 | 193745 |
| DDoS | 47 | 128014 |
| PortScan | 47 | 90694 |
| FTP-Patator | 47 | 5931 |
| SSH-Patator | 47 | 3219 |
| Web Attack | 47 | 2143 |
| Botnet | 47 | 1948 |
| Infiltration | 47 | 36 |
| Heartbleed | 47 | 11 |

### 4.3 Sampling methods

Despite the wide application areas, ADASYN sampling method has shortcomings. It is easily to be affected by outliers. If the K-nearest neighbors of a minority sample are all majority sample, its weight will become quite large, and more samples will be generated around it. In [19], a comprehensive method is proposed and had good results.

In this research, the same strategy was employed and the comprehensive method was used to balance the experimental datasets. ADASYN is used to oversample the training dataset to increase the number of Infiltration and Heartbleed these two attack behaviors data samples. Thus, the hyper parameters of oversampling methods are adjusted for the purpose of not synthesizing too many minority samples and reducing possible bad impact on the precision.

**4.4 Performance Evaluation Criteria**

In order to evaluate the performance of the classifier model, precision, recall and F-Measure, these most commonly used evaluation criteria in the field of intrusion detection prediction, are employed here to evaluate the proposed method. The three evaluation criteria are defined by confusion matrix, as shown in Table 4.

TN denotes the number of correct predictions, when a negative sample is considered as the negative, FP denotes the number of incorrect predictions, when a negative sample is considered as the positive. FN denotes the number of incorrect predictions, when a positive sample is considered as the negative. TP denotes the number of correct predictions, when a positive sample is considered as the positive. In this paper, the classification of network behaviors is a multi-classification problem, while the evaluation of it in confusion matrix using OvR (One Vs Rest) is the same as that of binary classification.

Table 4 Confusion matrix

| Predict Value / Actual Value | *Negative* | *Positive* |
|---|---|---|
| *Negative* | True Negative (TN) | False Positive (FP) |
| *Positive* | False Negative (FN) | True Positive (TP) |

*4.4.1 Precision*

Precision indicates how many of the samples that are predicted to be positive are correct, its value can be calculated by formula (2).

$$Precision = \frac{TP}{TP + FP} \quad (2)$$

*4.4.2 Recall*

Recall indicates how many positive samples in the sample that are correctly predicted. Its value can be calculated by formula (3).

$$Recall = \frac{TP}{TP + FN} \quad (3)$$

*4.4.3 F-Measure*

In order to evaluate the overall performance of the model and remove the effect of sample imbalance on the model, F-Measure is introduced to evaluate the experimental results. The values of F-Measure are shown by formula (4) with P stands for *Precision* and R stands for *Recall*.

$$F - Measure = \frac{(\beta^2 + 1) * P * R}{\beta^2 * P + R} \quad (4)$$

In the formula (4), $\beta$ is the adjustment parameter, and $\beta = 1$ is selected in the experiment.

At this time, F1-Score is shown by formula (5).

$$F1 - Score = \frac{2 * P * R}{P + R} \quad (5)$$

In order to compute metric within each class and average resulting metrics across classes, Macro-average of precision, recall and F1-Score was used as final criteria to evaluate the classification of various attacks behaviors. For Macro-average, each class has equal weight and will not be easily affected by majority classes like Micro-average, in which each instance has equal weight, and categories with small sample sizes are ignored.

*4.4.4 AUC*

AUC (area under curve) is defined as the area under the ROC and surrounded by the coordinate axis. ROC curve is a curve reflecting the relationship between false alarm rate and sensitivity. The x-axis is the false positive rate (FPR), and the closer the x-axis is to zero, the higher the precise is; the Y-axis is called the tire positive rate (TPR), and the larger the y-axis is, the better the accuracy is. According to the position of the curve, the whole graph is divided into two parts. The area under the curve is called AUC (area under curve), which is used to indicate the performance of prediction. The higher the AUC is, the larger the area under the curve is, the higher the performance of prediction is. The calculation formulas of FPR and TPR are shown in formulas 6 and 7.

$$FPR = \frac{FP}{TN + FP} \quad (6)$$

$$TPR = \frac{TP}{TP + FN} \quad (7)$$

The AUC value ranges from 0.5 to 1. The larger the AUC value is, the better the performance of the prediction model is. In this paper, One-Vs-Rest algorithm is also used to computes the average of the ROC AUC scores for each class against all other classes, then returns a uniformly-weighted average AUC.

**4.5 Experimental and Results Analysis**

In order to verify whether the sampling method mentioned and whether the ADASYN-Random Forest algorithm in this research can show good prediction performance, the comparison experiments are conducted. Three different sampling methods (RUS, SMOTE and ADASYN) combined with Random Forest are formed and conducted on CICIDS 2017, the experiments results are described in Table 5.

During the experiment, the oversampling strategy is set as (Infiltration: 500, Heartbleed: 500) based on the result of 5-fold cross-validation for 10 times, the under-sampling strategy for Random Under-Sampling is set as (Benign: 200000). The maximum depth of decision tree and number of estimators are 40 and 10, respectively. Whole experiments were carried out for 50 times respectively, and the average value was taken to obtain the final results (shown in Table 5 and Figure 3). Among them, the Random Forest algorithm combined with three different sampling methods proves that the ADASYN+ Random Forest method is the best.

Table 5 Comparison table of Random Forest combined with three sampling methods

| Combined Algorithm | macro-Precision | macro-Recall | macro-F1 | AUC |
|---|---|---|---|---|
| Random Forest | 98.421% | 90.460% | 93.409% | 0.995 21 |
| RUS + Random Forest | 95.463% | **93.606%** | 94.526% | 0.996 59 |
| SMOTE + Random Forest | 98.396% | 91.729% | 94.945% | 0.997 43 |
| ADASYN + Random Forest | **98.505%** | 92.303% | **95.303%** | **0.997 80** |

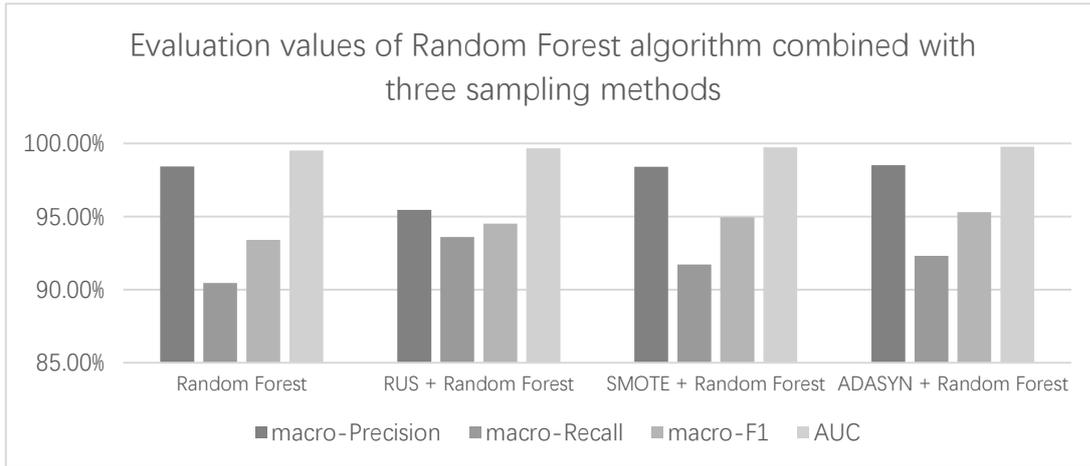

Figure 3. Evaluation values of Random Forest combined with three sampling methods

From the intrusion detection results of the CICIDS 2017 presented in Table 5, it can be seen that compared with the intrusion detection model based on Random Forest, the proposed ADASYN-Random Forest algorithm has better detection effect. Taking the macro-F1 as an example, the macro-F1 obtained by only Random Forest, RUS-Random Forest and SMOTE-Random Forest is 93.409%, 94.526% and 94.945% respectively, while the value obtained by ADASYN-Random Forest is the highest as 95.303%.

The macro-Precision values of these algorithms have no significant difference except for Random Forest with RUS, the values of rest three algorithms are 98.421% for only Random Forest, 98.396% for Random Forest with SMOTE and 98.505% for Random Forest with ADASYN. But Random Forest with Random Under-Sampling get the lowest macro- Precision value as 95.463%.

According to the results shown in the Figure 3, Random under sampling combined with Random Forest has the highest macro-Recall value as 93.606%, which is greater than those by using only Random Forest (90.460%), Random Forest with SMOTE (91.729%) and Random Forest with ADASYN (92.303%). The macro-Recall value acquired by ADASYN-Random Forest is not the highest but passable.

The performance of the algorithm is also evaluated according to AUC. For ADASYN-Random Forest, AUC = 0.997 80, and the optimal AUC is obtained comparing with those of other sampling methods.

## 5 CONCLUSION AND FUTURE WORK

This paper innovatively applies ADASYN-Random Forest to classify the network attack behaviors and tests the obtained model on a real unbalanced dataset, which achieves good prediction results. Compared with the traditional machine learning methods and random forest algorithm with other sampling methods, this method has higher prediction performance, efficiency and robustness. However, the parameters of ADASYN were artificially altered, more researches should be adopted in the future to find the most suitable sampling strategies. And for other under-sampling methods based on Clustering, if their computational complexity can be optimized and the parameters of under-sampling are controllable, they can be applied to imbalanced learning in the field of Intrusion Detection.

In addition, the dataset would be further optimized by capturing the network packet or dataflow[20] to improve the detection performance and a real network intrusion detector would be developed in practical application in the future work.

## ACKNOWLEDGMENTS

This work is supported by the Fund of Student Start-up Research of Zhejiang Normal University. Besides, the authors would like to thank the editors and anonymous reviewers for their constructive comments and suggestions for the improvement of this paper.